\documentclass[conference]{IEEEtran}
\IEEEoverridecommandlockouts
\usepackage{cite}
\usepackage{amsmath,amssymb,amsfonts}
\usepackage{algorithmicx}
\usepackage{graphicx}
\usepackage{textcomp}
\usepackage{xcolor}
\usepackage{algorithm}
\usepackage{algpseudocode}
\usepackage{multirow}
\usepackage{booktabs}
\usepackage{cleveref}
\usepackage{caption}
\usepackage{subcaption}
\usepackage[bottom]{footmisc}
\usepackage[T1]{fontenc}
\usepackage[utf8]{inputenc}
\usepackage{authblk}

\makeatletter
\newcommand\fs@spaceruled{\def\@fs@cfont{\bfseries}\let\@fs@capt\floatc@ruled
	\def\@fs@pre{\vspace{0.5\baselineskip}\hrule height.8pt depth0pt \kern2pt}%
	\def\@fs@post{\kern2pt\hrule\relax}%
	\def\@fs@mid{\kern2pt\hrule\kern2pt}%
	\let\@fs@iftopcapt\iftrue}
\makeatother

\def\BibTeX{{\rm B\kern-.05em{\sc i\kern-.025em b}\kern-.08em
		T\kern-.1667em\lower.7ex\hbox{E}\kern-.125emX}}

\begin{document}
	
	\title{SplitAMC: Split Learning for \\Robust Automatic Modulation Classification}
	
	\author[1]{Jihoon Park}
	\author[1]{Seungeun Oh}
	\author[2]{Seong-Lyun Kim}
	\affil[ ]{\small Department of Electrical and Electronic Engineering, Yonsei University, South Korea}
	\affil{\small \{jhpark, seoh\}@ramo.yonsei.ac.kr}
	\affil[2]{slkim@yonsei.ac.kr}

	\maketitle
	
	\begin{abstract}
		Automatic modulation classification (AMC) is a technology that identifies a modulation scheme without prior signal information and plays a vital role in various applications, including cognitive radio and link adaptation. With the development of deep learning (DL), DL-based AMC methods have emerged, while most of them focus on reducing computational complexity in a centralized structure. This centralized learning-based AMC (CentAMC) violates data privacy in the aspect of direct transmission of client-side raw data. Federated learning-based AMC (FedeAMC) can bypass this issue by exchanging model parameters, but causes large resultant latency and client-side computational load. Moreover, both CentAMC and FedeAMC are vulnerable to large-scale noise occured in the wireless channel between the client and the server. To this end, we develop a novel AMC method based on a split learning (SL) framework, coined \textit{SplitAMC}, that can achieve high accuracy even in poor channel conditions, while guaranteeing data privacy and low latency. In SplitAMC, each client can benefit from data privacy leakage by exchanging smashed data and its gradient instead of raw data, and has robustness to noise with the help of high scale of smashed data. Numerical evaluations validate that SplitAMC outperforms CentAMC and FedeAMC in terms of accuracy for all SNRs as well as latency.
	\end{abstract}
	
	
	\section{Introduction}
	With the remarkable development of wireless communication, understanding the radio spectrum plays an essential role in various applications, such as cognitive radio and link adaptation~\cite{1, 2}. In this respect, automatic modulation classification (AMC) is emerging as a promising technology in a way that the receiver (Rx) identifies the modulation scheme of the corresponding transmitter (Tx) signal without prior information~\cite{5}. As the first of its kind, a traditional AMC method aims to achieve high detection probability through the likelihood function design \cite{hameed2009likelihood}. However, due to its sensitivity to signal errors, it suffers from performance degradation and computational complexity, especially in the wireless environment where channel state information (CSI) is not given and channel gain fluctuates.
	
	\begin{figure}[htbp]	
		\begin{center}		
			\includegraphics[width = 1.0\linewidth]{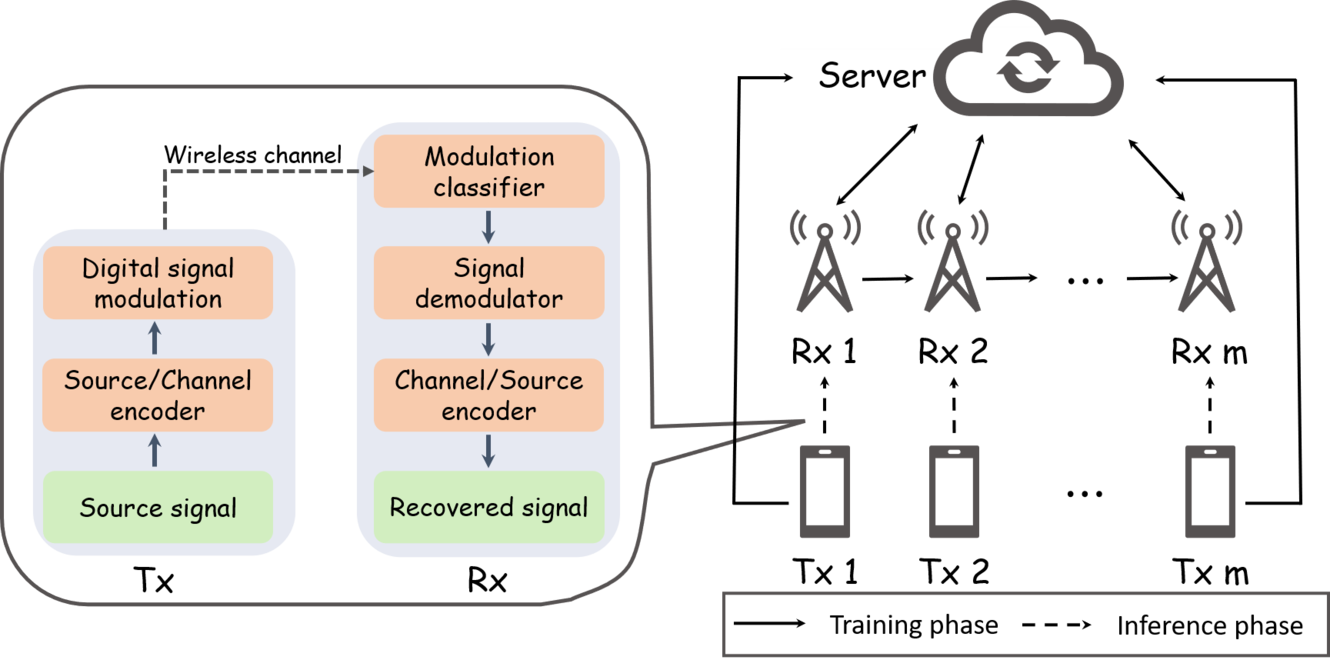}		
		\end{center}	
		\caption{An overview of communication system composed of Tx, Rx with AMC, and server.}
		\label{Fig:1}
	\end{figure} 
	
	The feature-based AMC method can detour these problems by extracting hand-crafted features from the received signal and performing classification tasks. Going further, DL-based AMC
	can solve the computational complexity problem of likelihood-based AMC under a time-varying channel with deep learning (DL) framework~\cite{4}. The existing DL-based AMC method is often rooted in a centralized architecture that enjoys dispersed data among multiple clients by direct local data aggregation on the server~\cite{10}. However, such centralized learning-based AMC (CentAMC) violates data privacy leakage while causing significant communication bottlenecks on the server-side. In order to guarantee data privacy, several AMC works apply a distributed learning framework to the AMC method. As its representative, FedeAMC~\cite{9}, which combines federated learning (FL) with AMC, performs modulation classification by updating the local model through aggregation and redistribution of the model parameters between the server and clients.
	However, this model exchange incurs a huge communication overhead, so it is ill-suited for the large-sized model. \emph{In addition, CentAMC and FedeAMC tend to be vulnerable to large-scale noise, leading to accuracy drop at low signal-to-noise ratio (SNR), highlighting the need for alternatives (see Table \ref{table:1}).}
	
	Towards a noise-robust, communication-efficient, and data-private AMC, this paper proposes a novel AMC method based on a split learning (SL) framework called \textit{SplitAMC}. The SplitAMC divides the entire deep neural network (DNN) into two partitions depending on cut-layer, an upper model segment and a lower model segment, each stored by the server and clients, respectively. Under this model-split architecture, clients and the server communicate the cut-layer representations, so-called \textit{smashed data}, and its corresponding gradients. In doing so, the SplitAMC can benefit from high accuracy over large-scale noise, thanks to the scale of smashed data shown in Fig.~\ref{Fig:3}. Also, exchanging smashed data instead of model parameters or raw data results in improved communication efficiency as well as data privacy guarantee of SplitAMC.
	
	The contributions of this paper are summarized below:
	\begin{itemize}
		\item By revisiting the SL framework~\cite{gupta2020secure}, we propose SplitAMC, an AMC method based on SL, that enables smashed data exchange instead of model parameters or raw data exchange.
		
		\item SplitAMC's smashed data exchange solves data privacy leakage that occurs in 
		CentAMC, while reducing latency at the same time. Latency analysis for SplitAMC is available in Sec. IV.
		
		\item \textit{Thanks to the large scale of the smashed data, SplitAMC has robust accuracy even in a large-noise environment, which is proven by experiments.}
	\end{itemize}


	The rest of the paper is organized as follows: Sec.~II introduces a system model, including a single-carrier system-based signal model and a wireless channel model. Next, Sec.~III describes the operation of our proposed SplitAMC in detail. In Sec. IV, the performance of SplitAMC is validated and analyzed through extensive simulations, compared to CentAMC as well as FedeAMC.

	
	\section{System Model}
	In this section, we describe the network topology in which the DL-based approaches including the proposed SplitAMC run, and then sequentially describe the signal model and channel model of communication links included in the network.
	
	\subsection{Network Topology}
	As described in Fig. \ref{Fig:1}, we consider a communication system for AMC consisting of Tx-Rx pairs and server. 
	
	\textit{1) Tx-Rx Communication Link:} Each Tx passes the source signal through the sampler and quantizer to perform analog-to-digital conversion (ADC) and encodes it, followed by selecting the modulation type, controlling the transmit power, and transmitting it to the corresponding Rx. Then, each Rx infers the modulation type of Tx through a modulation classifier and demodulates it based on it. Consequently, the goal of this paper is to train a modulation classifier with high accuracy, the key to the aforementioned communication system, via DL-based approach. 
	
	\textit{2) Rx-Server Communication Link:} To achieve this goal, DL-based approaches perform additional communication between Rx and a server to train the model. 
	At this time, an analog-modulated signal is assumed to be used. Although digital communication is widely used, it has notable drawbacks such as significant performance drop when deviating from target channel conditions (i.e., outage due to frequent channel fluctuations). On the other hand, analog communication can avoid this and even improve performance thanks to the regularizer effect of the noise \cite{zhou2021learning}. Furthermore, analog-modulated signals can enjoy enhanced data privacy as well as latency gains by enabling over-the-air computation (AirComp) \cite{koda2021airmixml}.

	\subsection{Communication Model}
	
	\subsubsection{Signal Model}
	
	For Tx-Rx communication link, we use a regular signal model of the unknown single-input single-output single-carrier systems as in \cite{11, 13}. Let $r(n)$ represent the unknown modulated signals at the Rx, denoted by:
	\begin{equation}
		\begin{split}
			& r(n) = A_ne^{j(2\pi f_0nT+\theta_n)}s(n) +\sigma(n), \\
			& n \in \{0, 1, \cdots, N-1\},
		\end{split}
	\end{equation}
	where $s(n)$ and $\sigma(n)$ are the transmitted modulation signal and additive white Gaussian noise (AWGN), respectively. In addition, $A_n$ is the signal amplitude of symbol $n$, following rayleigh channel fading, $f_0$ is the carrier frequency offset, $\theta_n$ is the time-varying carrier phase offset of symbol $n$, $T$ is the symbol interval, and $N$ is the number of symbols of the signals and also the number of samples to be used for training of the DL model.
	
	The IQ sample, which is the training and test sample of DL, consists of in-phase ($I$) and quadrature ($Q$) parts of $r(t)$. Based on the received signal model, the IQ sample can be expressed as: 
	\begin{equation}
		\begin{split}
			& I = \{Real(r(0)), Real(r(1)), \cdots\, Real(r(N-1))\}, \\
			& Q = \{Imag(r(0)), Imag(r(1)), \cdots\, Imag(r(N-1))\},
		\end{split}
		\label{eq:2}
	\end{equation}
	where $Real(\cdot)$ and $Imag(\cdot)$ are functions for extracting values corresponding to the real and imaginary parts of the received signal $r(t)$.
	
	Also, we can define the SNR of the received signal as follows:
	\begin{equation}
		\begin{split}
			\gamma_{data} = 10\cdot\log_{10} (\frac{\sum_{n=0}^{N-1}{|A_ns(n)|}^2}{\sum_{n=0}^{N-1}{|\sigma(n)|}^2}) [dB].
		\end{split}
	\end{equation}
	
	\begin{figure*} [t]
			\includegraphics[width=0.95\linewidth]{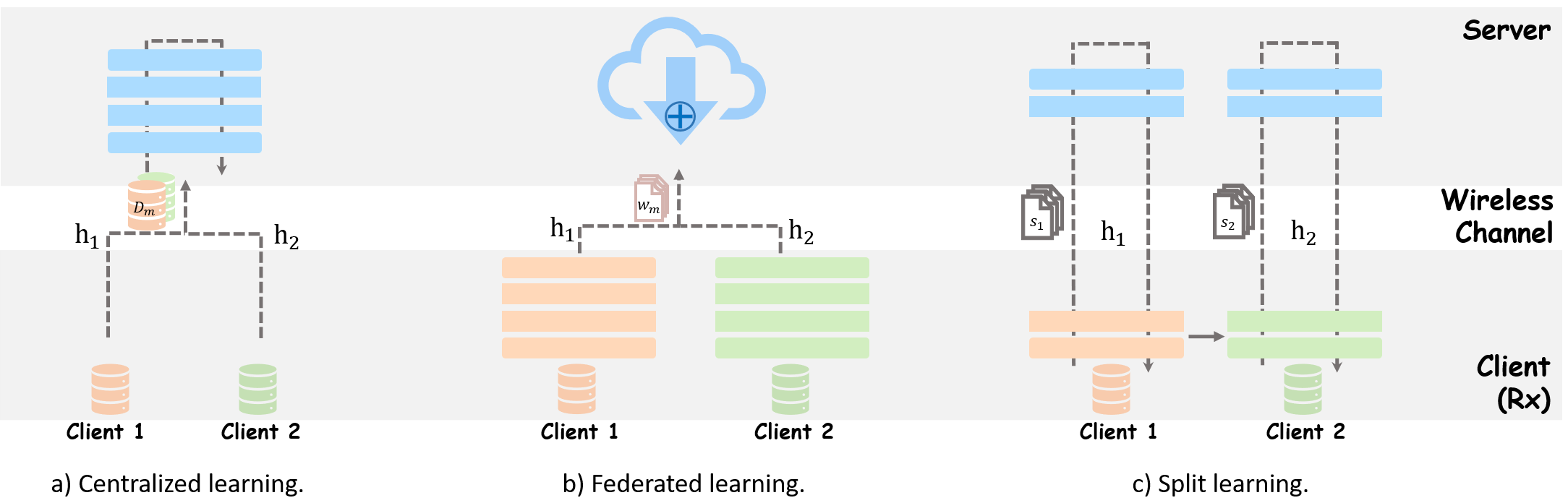}		
		\caption{Graphical illustrations of a) centralized learning, b) federated learning (FL), and c) split learning (SL).}	
		\label{Fig:2}
	\end{figure*}
	
	\subsubsection{Channel Model}
	For Rx-server communication link, we consider the path-loss attenuation, channel fading, and noise, and its SNR is expressed by the following equation:
	\begin{equation}
		\gamma = 10\cdot\log_{10} (\frac{h\cdot P\cdot d^{-\alpha}}{\sigma^2}) [dB],
		\label{eq:1}
	\end{equation}
	where $P$ is the transmit power, $d$ is the distance between Tx and Rx, $\alpha$ is the path loss attenuation exponent, which usually corresponds to a value greater than or equal to 2, and $\sigma^2$ is the variance value of the AWGN in the wireless channel environment. Moreover, $h$ indicates channel fluctuations and follows an exponential distribution with a mean of 1 ($h\sim{\exp{(1)}}$).

	\section{Split learning-based AMC}
	This section introduces the operation of SplitAMC method along with two existing AMC methods, CentAMC and FedeAMC.
	Let $m$ be the subscript for Rx. Then, $m$-th Rx of set $\mathbb{M}= \{1, 2, \cdots, M\}$ produces the IQ data samples via \eqref{eq:2}. For all $m\in\mathbb{M}$, the local dataset $D_m = (x_m, y_m)$ consists of IQ samples $x_m$ and its corresponding ground-truth labels $y_m$. As depicted in Fig. \ref{Fig:2}c, to enable model-split architecture shown in \cite{gupta2020secure}, the entire model weight $w_m$ is divided into the upper model segment $w_s$ and the lower model segment $w_{c,m}$ based on the cut-layer.
	
	\subsection{Training Phase: The Operation of SplitAMC}
	
	
	\noindent\textbf{Client-side Forward Propagation (FP).} \quad The $m$-th Rx randomly selects $B$ data-label tuples from the local dataset $D_m$ to compose a batch, and passes it through the lower model segment $w_{c,m}$ to generate smashed data $s_m$ as follows:
	\begin{equation}
		s_m = f(x_m;w_{c,m}), 
	\end{equation}
	where $f$($\cdot$) denotes a function that maps $x_m$ to $s_m$, determined by $w_{c,m}$.
	Then, the $m$-th Rx sends $s_m$ to the server through an analog-modulated signal. $\bar{s}_m$ reflecting the fluctuation and noise of the uplink (UL) wireless channel between Rx and the server in $s_m$ is expressed as belows:
	\begin{equation} \label{eqn:smashed data}
		\bar{s}_m = h\cdot s_m + \mathbb{N} = h\cdot f(x;w_{c,m}) + \mathbb{N},
	\end{equation}
	where $\mathbb{N}$ follows a zero-mean complex Gaussian distribution with variance $\sigma^2$.
	
	\noindent\textbf{Server-side FP \& Backpropagation (BP).} \quad To update the server-side model, the smashed data with Gaussian noise $\bar{s}_m$ becomes the input of the upper model segment located on the server, yielding a \textit{softmax output} $\hat{y}_m$ as follows:
	\begin{equation}
		\hat{y}_m = g(\bar{s}_m;w_s),
	\end{equation}
	where $g$($\cdot$) is a function that maps $\bar{s}_m$ to $\hat{y}_m$ due to the upper model segment $w_s$.
	
	Then, by using the cross-entropy function, the loss $L_{CE}$ can calculated by:
	\begin{equation}
		L_{CE} = -\frac{1}{d_y}\sum_{i=1}^{d_y} y_{m,i} \log(\hat{y}_{m,i}),
		\label{eq:3}
	\end{equation}
	where $i$ is the subscript for element, so that $y_{m,i}$ and $\hat{y}_{m,i}$ are $i$-th elements of ground-truth label $y_{m}$ and prediction $\hat{y}_{m}$, respectively, while $d_y$ denotes the dimension of $y_{m,i}$. 
	
	With the aid of downlink (DL) communication in the cut-layer, BP is available allowing model update of Rx and server as follows:
	\begin{equation}
		\left[
		\begin{matrix}
			w_{c,m}\\
			w_s
		\end{matrix}
		\right] \leftarrow 
		\left[
		\begin{matrix}
			w_{c,m}\\
			w_{s}
		\end{matrix}
		\right] - \eta \left[
		\begin{matrix}
			\nabla_{w_{c,m}} L_{CE} \\
			\nabla_{w_{s}} L_{CE}
		\end{matrix}
		\right],
	\end{equation}
	where $\eta$ and $\nabla_{w_{c,m}(w_s)}$ is the learning rate and the partial derivative with respect to $w_{c,m}(w_s)$, respectively.
	
	After that, the $m$-th Rx transmits its lower model segment to the $(m+1)$-th Rx as follows:
	\begin{equation}
		w_{c,m+1}\leftarrow{w_{c,m}},
	\end{equation}
	completing the single communication round of SplitAMC. The training operation of SplitAMC is detailed in the pseudocode of \textbf{Algorithm 1}.
	
	\subsection{Inference Phase: Performance Metrics}
	When the model $w_m=[w_{c,m},w_s]$ converges after the $K$ communication rounds, it enters the inference phase. To this end, two types of inference method can be considered. In the first case, all clients download the shared upper model segment of the server, then use it as a classifier for the modulated signal transmitted from the paired Tx, followed by demodulation. In other cases, we can consider an inference method in which the client and server communicate the smashed data of test data and prediction through UL and DL communication, respectively, while retaining the model.
	In both inference methods, it is possible to measure the performance of the test model $w_m$ via the following performance metrics:
	
	\subsubsection{Classification Performance}
	
	The correct classification probability $P_{cc}$ is employed to evaluate how accurately the test model $w_m$ classifies the modulation scheme. When the total number of test samples is $N_{test}$, $P_{cc}$ is defined by :
	\begin{equation}
		P_{cc} = \frac{N_{correct}}{N_{test}} \times 100 [\%],
	\end{equation}
	where $N_{correct}$ is the number of samples that successfully classified the modulation scheme among all $N_{test}$ test samples.
	
	\floatstyle{spaceruled}
	\restylefloat{algorithm}
	\begin{algorithm}[t]{
			\caption{Model training method of SplitAMC framework.}
			\begin{algorithmic}[1]
				\State \textbf{requirements:} $k=0$, $w=[w_{c,m},w_s]$, $D_m$ $\forall m$\\
				Hyperparameters:
				\begin{itemize}
					\item number of clients: $|\mathbb{M}|=M$,
					\item total number of communication rounds: $K$,
					\item learning rate: $\eta$				
				\end{itemize}
				\While{$k<K$}
				\State \hspace{-15pt} Client $m\in \mathbb{M}$:  
				\State \hspace{-10pt} \textbf{produces} $s_m$ via (5) \emph{\hfill$\triangleright$ Client-side FP}
				\State \hspace{-10pt} \textbf{unicasts} $(s_{m}, y_{m})$ to the server	\emph{\hfill$\triangleright$ UL Communication}	
				\State \hspace{-15pt} Server:
				\State \hspace{-10pt} \textbf{produces} $\hat{y}_m$ via (7) \emph{\hfill$\triangleright$ Server-side FP}
				\State \hspace{-10pt} \textbf{produces} $L_{CE}$ via (8) 
				\State \hspace{-10pt} \textbf{updates} $w_s$ via (9) \emph{\hfill$\triangleright$ Server-side BP}
				\State \hspace{-10pt} \textbf{unicasts} $\nabla_{\bar{s}_m} L_{CE}$ \emph{\hfill$\triangleright$ DL Communication}

				\State \hspace{-15pt} Client $m\in \mathbb{M}$:  
				\State \hspace{-10pt} \textbf{updates} $w_{c,m}$ via (9) \emph{\hfill$\triangleright$ Client-side BP}
				\State \hspace{-10pt} \textbf{unicasts} $w_{c,m}$ to $(m+1)$-th Rx as in (10) \emph{\hfill$\triangleright$ Model Transfer}			
				\State \hspace{-15pt} $k\leftarrow{k+1}$
				\EndWhile
			\end{algorithmic}     
	}\end{algorithm}
	
	\subsubsection{Latency Model}
	
	Meanwhile, we can model the latency that occurs in the second inference method mentioned above or training phase. We divide the overall latency into communication latency $T_{comm}$, consisting of UL latency $T_{UL}$ \& DL latency $T_{DL}$, and computation latency $T_{comp}$. Here, we assume a static channel condition, i.e., $h=1$ in \eqref{eq:1} for convenience. 
	
	In terms of communication latency, it is proportional to the number of transmitted bits and inversely proportional to the channel capacity, under the assumption of a static channel. Let $L_a^b$ and $\beta_a^b$ ($a\in{\{SL, CL, FL}\}$, $b\in{\{UL, DL}\}$) denote the number of parameters exchanged and the number of bits per parameter during $b$ communication between Rx and server in $a$-based AMC method, respectively. Then, for the UL (DL) transmission rate $R_{UL(DL)}=BW\cdot log_2({1+\gamma})$ with $\gamma$ in \eqref{eq:1} where $h=1$ and channel bandwidth $BW$, the latency for UL (DL) communication is $\frac{L_a^{UL(DL)}\cdot{\beta_a^{UL(DL)}}}{R_{UL(DL)}}$ for all $a$ method.
	
	Computational latency $T_{comp}$ is classified into client-side latency $(T_{client})$ and server-side latency $(T_{server})$. Both latencies are expressed as the ratio of $C$ to $f$, where $C$ is the number of CPU cycles for processing FP as well as BP, and $f$ is for client-side or server-side computational capacity (in CPU cycles$/$s). Assuming that the server-side computational capacity is infinite, $T_{client}$ and $T_{server}$ become unit computing time $\tau_{comp}$ and 0, respectively, during 1 communication round.
	
	Table I summarizes $T_{comm}$ and $T_{comp}$ for 1 communication round of SplitAMC as well as FedeAMC and CentAMC. Note that the proposed SplitAMC can benefit in terms of computational latency, multiplied by the $\lambda \in[0, 1]$, depending on where the cut-layer is located.
	
	\subsection{Other AMC Methods}
	
	\subsubsection{CentAMC~\cite{10}} As described in Fig.~2a, in CentAMC, all local datasets $D_m$ are aggregated on the server through a wireless channel to form a single global dataset, becoming an input for a global model training of the server. By doing this, CentAMC easily obtains data diversity gain, but it causes a communication bottleneck when sending raw samples and violates data privacy guarantees.
	
	\subsubsection{FedeAMC~\cite{9}}
	In FedeAMC, the server distributes the global model to each local client $m\in\mathbb{M}$. Then, all local clients train the model using local dataset $D_m$. After that, as shown in Fig. \ref{Fig:2}b, local model parameters $w_{m}$ are transmitted to the server through the wireless channel link. The server takes weighted averaging of aggregated local parameters, yielding a global model $w$, i.e., $w=(\sum_{m=1}^{M} {|D_m|\cdot{w_m}})/\sum_{m=1}^{M}{|D_m|}$. This approach can take data diversity gain without direct data exchange, but large communication 
	overhead occurs for large-sized models.
	
	\begin{table}[t] 
		\setlength{\tabcolsep}{0.4cm}
		\renewcommand{\arraystretch}{2.0}
		\caption{A summary of communication and computation latency for a single communication round of AMC methods.}
		\label{table:2}
		\begin{center}
			\begin{tabular}{|c|cc|c|}
				\toprule[1pt] 
				\hline
				\multirow{2.3}{*}{\textit{Latency}} & \multicolumn{2}{c|}{$T_{comm}$} & \multicolumn{1}{c|}{$T_{comp}$} \\
				\cline{2-4}
				& $T_{UL}$ & \multicolumn{1}{c|}{$T_{DL}$} & \multicolumn{1}{c|}{$T_{client}$} \\
				\hline \bottomrule
				\textbf{SplitAMC} & $\frac{L_{SL}^{UL} \cdot \beta_{SL}^{UL}}{R_{UL}}$ & $\frac{L_{SL}^{DL} \cdot \beta_{SL}^{DL}}{R_{DL}}$ & $\lambda \cdot \tau_{comp} $ \\
				\textbf{FedeAMC} & $\frac{L_{FL}^{UL} \cdot \beta_{FL}^{UL}}{R_{UL}}$ & $\frac{L_{FL}^{DL} \cdot \beta_{FL}^{DL}}{R_{DL}}$ & $ \tau_{comp}$ \\
				\textbf{CentAMC} &  $\frac{L_{CL}^{UL} \cdot \beta_{CL}^{UL}}{R_{UL}}$ & $\frac{L_{CL}^{DL} \cdot \beta_{CL}^{DL}}{R_{DL}}$  & - \\ \hline \bottomrule[1pt] 
			\end{tabular}
		\end{center}
	\end{table}
	
	\begin{table*}[t] 
		\setlength{\tabcolsep}{0.2cm}
		\renewcommand{\arraystretch}{1.5}
		\caption{Classification performance $P_{cc}$ of SplitAMC and comparison groups on datasets with different SNRs $\gamma_{data}$.}
		\label{table:1}
		\begin{center}
			\begin{tabular}{|ccccccccccccc|}
				\toprule[1pt] \hline
				\multirow{3.2}{*}{Channel SNR} & \multicolumn{4}{|c|}{$\gamma_{data} = $ 5dB} & \multicolumn{4}{|c|}{10dB} & \multicolumn{4}{|c|}{15dB}   \\
				\cline{2-13}
				& \multicolumn{2}{|c|}{\textbf{Avg. SNR ($= \gamma_{avg}$ )}} & \multicolumn{2}{|c|}{\textbf{Fixed SNR ($ = \gamma_{fix}$)} } & \multicolumn{2}{|c|}{\textbf{Avg. SNR}} & \multicolumn{2}{|c|}{\textbf{Fixed SNR }} & \multicolumn{2}{|c|}{\textbf{Avg. SNR}} & \multicolumn{2}{|c|}{\textbf{Fixed SNR }}
				\\ \cline{2-13}
				& \multicolumn{1}{|c}{-10dB} &\multicolumn{1}{c|}{+10dB} &\multicolumn{1}{|c}{-10dB} &\multicolumn{1}{c|}{+10dB} & \multicolumn{1}{|c}{-10dB} &\multicolumn{1}{c|}{+10dB} &\multicolumn{1}{|c}{-10dB} &\multicolumn{1}{c|}{+10dB} & \multicolumn{1}{|c}{-10dB} &\multicolumn{1}{c|}{+10dB} &\multicolumn{1}{|c}{-10dB} &\multicolumn{1}{c|}{+10dB} 
				\\ \cline{1-13}\bottomrule
				\textbf{SplitAMC$_{(1,3)}$}          & \multicolumn{1}{|c}{81.2 \%}       &  \multicolumn{1}{c}{82.3 \%}      & \multicolumn{1}{|c}{81.7 \%}       &   \multicolumn{1}{c|}{82.0 \%}     & \multicolumn{1}{|c}{97.8 \%}       &  \multicolumn{1}{c}{98.6 \%}      & \multicolumn{1}{|c}{98.6 \%}       &   \multicolumn{1}{c|}{99.0 \%}       & \multicolumn{1}{|c}{99.6 \%}       &  \multicolumn{1}{c}{99.9 \%}      & \multicolumn{1}{|c}{99.7 \%}       &   \multicolumn{1}{c|}{99.9 \%}	
				\\ \hline
				\textbf{SplitAMC$_{(2,2)}$}          & \multicolumn{1}{|c}{81.1 \%}       &  \multicolumn{1}{c}{81.1 \%}      & \multicolumn{1}{|c}{80.8 \%}       &   \multicolumn{1}{c|}{82.5 \%}      & \multicolumn{1}{|c}{97.7 \%}       &  \multicolumn{1}{c}{98.9 \%}      & \multicolumn{1}{|c}{98.6 \%}       &   \multicolumn{1}{c|}{99.0 \%}  & \multicolumn{1}{|c}{99.6 \%}       &  \multicolumn{1}{c}{99.9 \%}      & \multicolumn{1}{|c}{99.8 \%}       &   \multicolumn{1}{c|}{99.9 \%}
				\\ \hline
				\textbf{CentAMC}           & \multicolumn{1}{|c}{66.6 \%}       &  \multicolumn{1}{c}{67.5 \%}      & \multicolumn{1}{|c}{68.4 \%}       &   \multicolumn{1}{c|}{68.7 \%}   & \multicolumn{1}{|c}{77.3 \%}       &  \multicolumn{1}{c}{78.4 \%}      & \multicolumn{1}{|c}{77.9 \%}       &   \multicolumn{1}{c|}{78.7 \%}   & \multicolumn{1}{|c}{97.4 \%}       &  \multicolumn{1}{c}{97.5 \%}      & \multicolumn{1}{|c}{97.7 \%}       &   \multicolumn{1}{c|}{97.8 \%}
				\\ \hline
				\textbf{FedeAMC}           & \multicolumn{1}{|c}{64.7 \%}       &  \multicolumn{1}{c}{64.4 \%}      & \multicolumn{1}{|c}{64.4 \%}       &   \multicolumn{1}{c|}{65.1 \%}    & \multicolumn{1}{|c}{79.4 \%}       &  \multicolumn{1}{c}{79.2 \%}      & \multicolumn{1}{|c}{79.1 \%}       &   \multicolumn{1}{c|}{78.6 \%}  & \multicolumn{1}{|c}{93.0 \%}       &  \multicolumn{1}{c}{93.5 \%}      & \multicolumn{1}{|c}{92.9 \%}       &   \multicolumn{1}{c|}{93.4 \%}
				\\ \hline \bottomrule[1pt] 
			\end{tabular}
		\end{center}
	\end{table*}

	\section{Experimental Results}
	This section evaluates SplitAMC's performance in terms of accuracy, convergence speed, and latency. For comparison, we use the aforementioned CentAMC and FedeAMC, and also SplitAMC with different cut-layer location. For the ResNet-18 model \cite{18} with 4 residual blocks, $\text{SplitAMC}_{(1,3)}$ and $\text{SplitAMC}_{(2,2)}$ represent the SplitAMC framework when the cut-layers are located after the 1st and 2nd blocks, respectively.
	
	For signal generation between Tx and Rx, Communications Toolbox in MATLAB is adopted. Considering the modulation schemes of QPSK, 16-QAM, and 64-QAM when SNR $\gamma_{data}$ is 5, 10, and 15 dB, each modulation scheme includes 5000 data symbols, each composed of 1000 IQ samples. Prior to passing the CNN model, each IQ sample is transformed into a constellation image and resized to 64$\times$64 dimensions as in \cite{17, 18}. For experiment, we employ Intel i7-10700 CPU and GTX 3080Ti GPU, where the software environment is based on PyTorch v1.10.0 and CUDA v11.3 with Python v3.9.4. Other simulation parameters are given as: $M=2$, batch size $B=64$, total communication rounds $K=50000$ steps, learning rate $\eta=0.004$, transmit power $P=100$ mW, distance between Rx-server $d = 100$ m, and path-loss exponent $\alpha = 2$.
	
	\subsection{Accuracy \& Convergence Speed}
	
\begin{figure} [t]
		\begin{subfigure}{1.0\linewidth}
			\centering
			\includegraphics[width=0.9\linewidth]{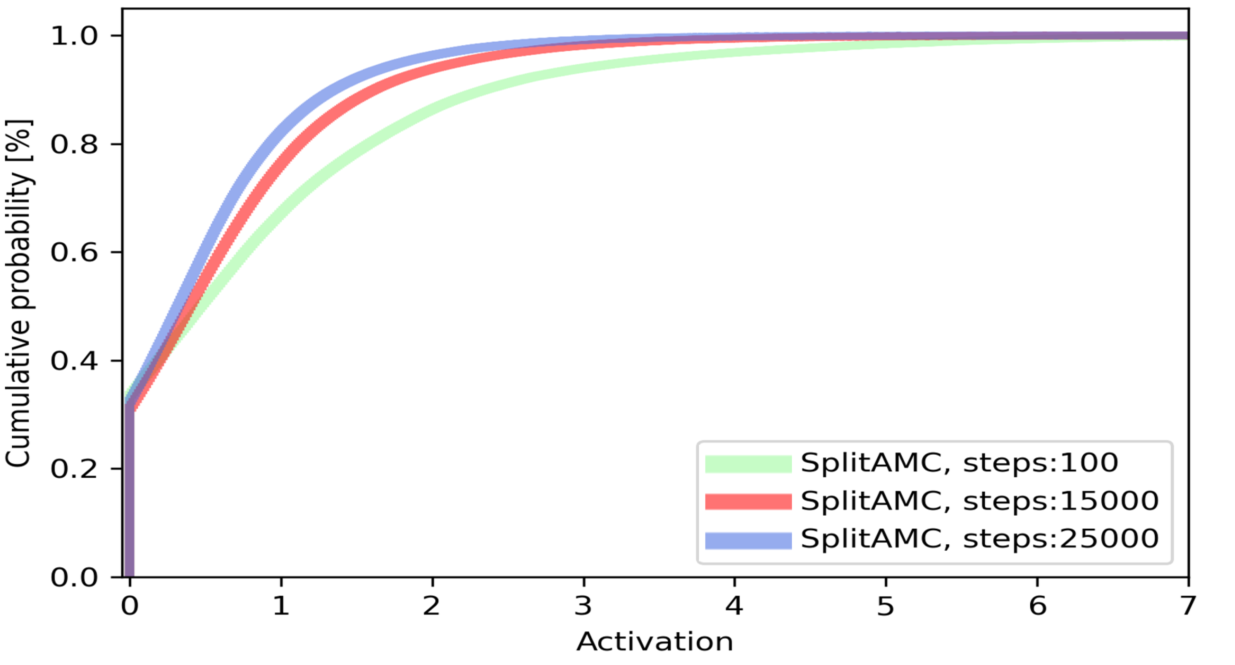}
			\caption{The CDF of $s_m$.}
		\end{subfigure}
		\hfill
		
		\begin{subfigure}{1.0\linewidth}
			\centering
			\includegraphics[width=0.9\linewidth]{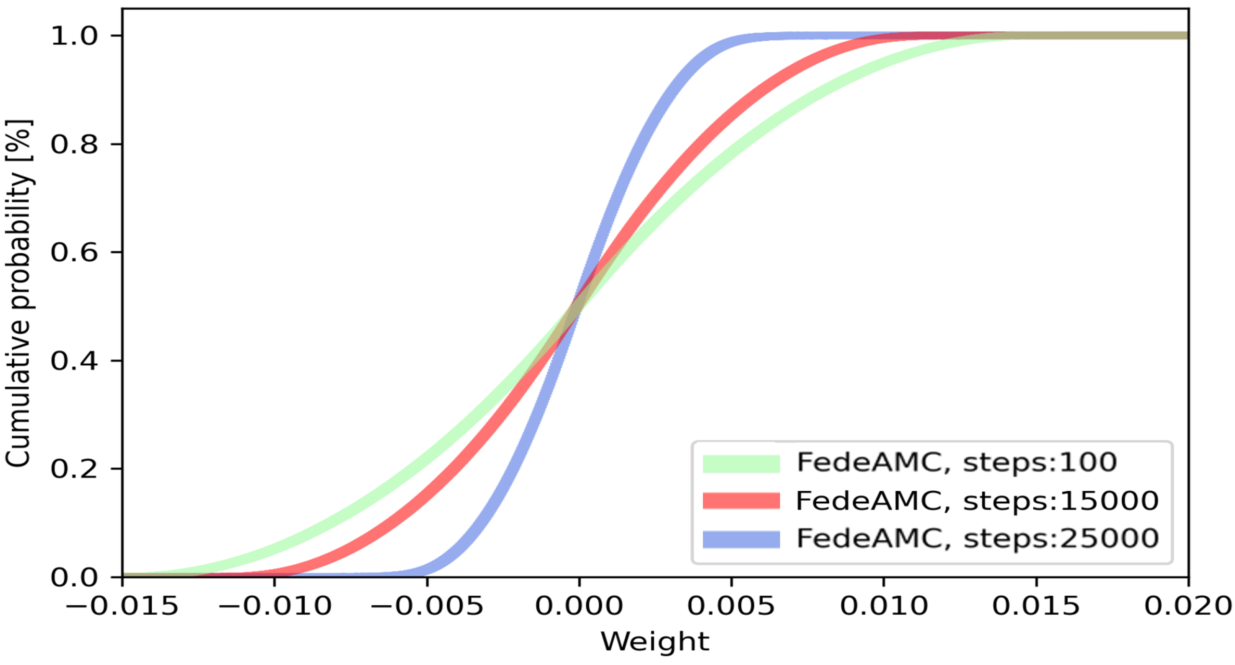}
			\caption{The CDF of $w_m$.}
		\end{subfigure}
		\caption{Cumulative distributions of parameter values in SplitAMC and FedeAMC for different steps.}	
		\label{Fig:3}
	\end{figure}

\begin{figure} [t]
		\begin{subfigure}{1.0\linewidth}
			\centering
			\includegraphics[width=0.75\linewidth]{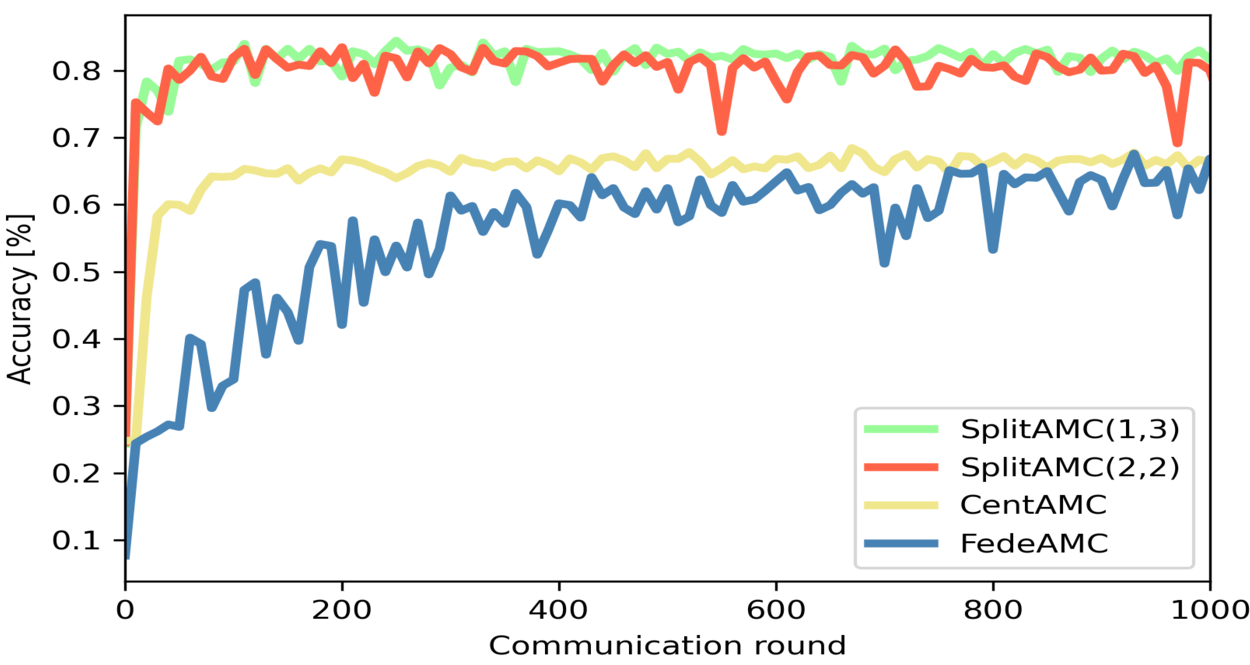}
			\caption{Learning curves when $\gamma_{data}=5$ dB.}
		\end{subfigure}
		\hfill
		
		\begin{subfigure}{1.0\linewidth}
			\centering
			\includegraphics[width=0.75\linewidth]{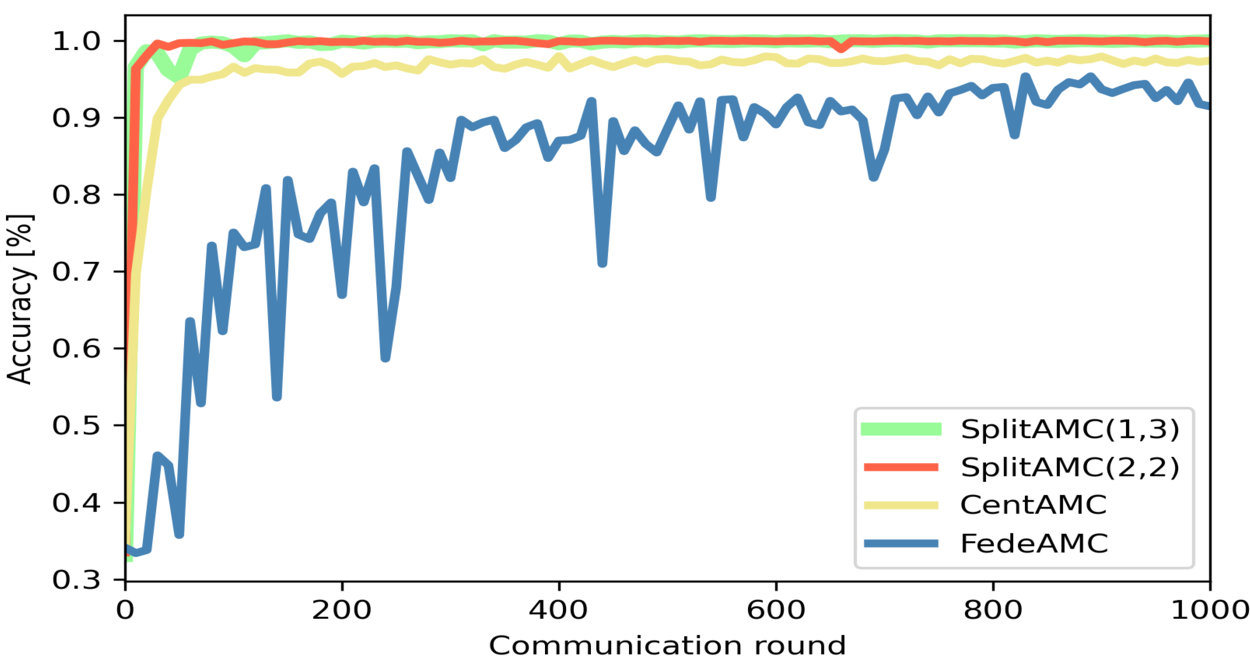}
			\caption{Learning curves when $\gamma_{data}=15$ dB.}
		\end{subfigure}
		\caption{Learning curves of AMC methods w.r.t different $\gamma_{data}$.}
		\label{Fig:4}	
	\end{figure}

	Table II shows the classification accuracy of AMC methods according to different $\gamma_{data}$ and channel environments. We consider two wireless communication environments depending on whether there is channel fluctuation, where the averaged SNR with channel fluctuation $\gamma_{avg}$ and the fixed SNR without channel fluctuation $\gamma_{fix}$ are the same as $\gamma$ with $h=1$ in (4). First, in all cases except for a few cases when $\gamma_{data}$ is 10dB, the classification accuracy 
	is high in the order of SplitAMC, CentAMC, and FedeAMC, showing the \textbf{noise-robustness} of SplitAMC. This is rooted in the scale difference of parameters exchanged through the Rx-server link in SplitAMC and FedeAMC shown in Fig. \ref{Fig:3}, respectively. Considering also CentAMC, which uploads raw data with mean and variance normalized to 0.5, when noise of the same size is injected, the relatively large parameter scale of SplitAMC leads to small performance degradation.
	These parameter scale gaps can be compensated by increasing the transmit power (e.g., 200 times the transmit power [mW] for  200 times scale difference) which is unachievable given the client's power constraints.
	Fig. \ref{Fig:3} also implies a faster convergence speed of SplitAMC compared to FedeAMC through the parameter distribution change according to the number of steps, which is confirmed from the learning curve of Fig. \ref{Fig:4}.
	
	Returning to Table II, the classification performance tends to deteriorate as $\gamma_{data}$ or $\gamma_{avg}(\gamma_{fix})$ decreases and when channel fluctuations exist. $\gamma_{data}$ and $\gamma_{avg}(\gamma_{fix})$ are respectively related to the noise inherent in the IQ data itself and the noise applied to the resized constellation image, and it is confirmed that the performance reduction for $\gamma_{data}$ is noticeable among them, indicating the importance of the Tx-Rx communication link state. The existence of channel fluctuations reduces the accuracy since it is difficult to learn a generalized model in the training phase. 
	Focusing on the gap in classification accuracy between AMC methods, the worse the communication channel conditions, the greater the difference, proving the effectiveness of SplitAMC. In addition, in Table II as well as in Fig. \ref{Fig:4}, there is no significant difference in performance in terms of accuracy or convergence speed for cut-layer in SplitAMC.
	
	\subsection{Latency Measurement for Training}
\begin{figure} [t]
		\centering
		\includegraphics[width=0.85\linewidth]{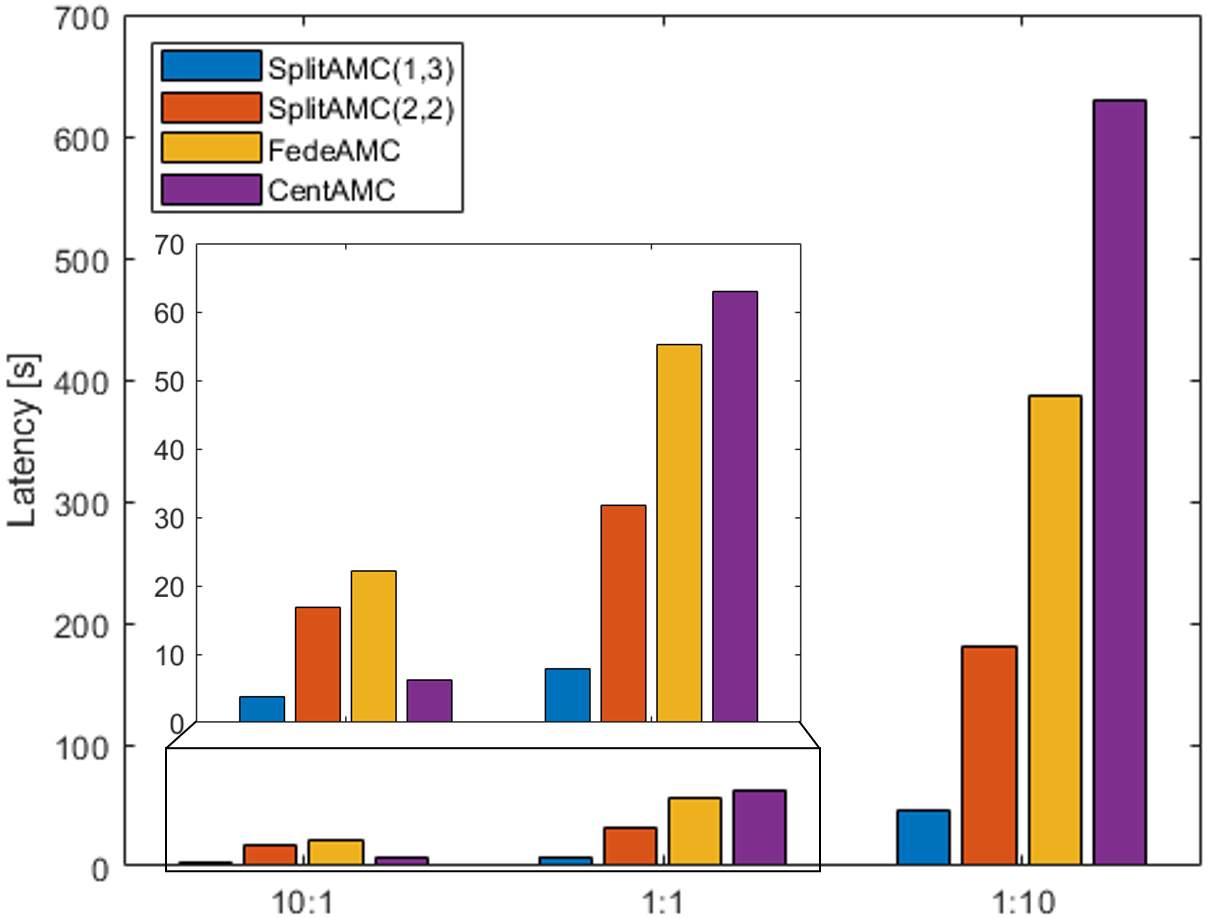}
		\caption{Latency according to the ratio of $\tau_{comp}$ to $\tau_{comm}$.}
		\hfill
		\label{Fig:5}
	\end{figure}

	Fig. \ref{Fig:5} compares latency between AMC methods according to the ratio of $\tau_{comp}$ and $\tau_{comm}$, which is the reciprocal of uplink rate $R_{UL}$. Here, to reflect the UL-DL asymmetric capacity of the cellular network, it is assumed that $R_{DL} = 10 R_{UL}$. Other parameters for calculating latency are as follows: $\beta_{a}^b=32$ bits for all $a$ \& $b$, $\gamma=10$ dB for UL communication, $BW=10$ MHz, and $50$ total communication rounds.
	
	The first thing to note is that overall latency performance improves when the ratio is 10:1, i.e., when $\tau_{comm}$ is small. In other words, this means that the communication payload size of UL, which determines UL latency, is the main factor influencing latency. Thanks to the low dimension of smashed data, SplitAMC outperforms the baselines, except for CentAMC at 10:1. In the case of CentAMC, $T_{comp}$ converges to 0 under our premise that the server-side computational capacity is infinite, and this dramatically reduces overall latency when $\tau_{comp}:\tau_{comm}$ is 10:1. Regarding the cut-layer, the closer the cut-layer is to the input layer, the better in terms of latency. This is because the client-side computational latency varies according to $\lambda$ as the cut-layer location changes, although there is no huge gap in the dimension of activation. When considering its accuracy and convergence speed together, it is optimal to place the cut-layer close to the input layer in SplitAMC, even beneficial in terms of client memory. 
	Note that the above method is also applicable for the inference phase, by replacing the payload size.


	\section{Conclusion}
	In this paper, we revisited the SL framework and applied it to AMC to design SplitAMC framework. Unlike the existing AMC methods, in SplitAMC, the client and server each hold a fraction of the entire model and exchange the output and gradient of the cut-layer, thereby resulting in a small memory size, computational cost, and communication payload size. Moreover, with the help of large scale inherent in smashed data, SplitAMC guarantees noise-robustness that 
	can achieve high accuracy even when noise with large variance is ejected. Numerical evaluations validate the accuracy, convergence speed, and latency of SplitAMC. It is worth noting that this work focused on connecting parametric properties in distributed learning (i.e., scale and dimension) to metrics in wireless communication (i.e., classification accuracy and latency). As a future work, we will exploit the classification performance as the number of clients increases as in~\cite{20}. Also, investigating the accuracy-privacy tradeoff for noise variance~\cite{oh2022differentially} in SplitAMC could be an interesting topic, deferred as future work.

	\section*{Acknowledgement}
	This work was supported by Institute of Information \& communications
	Technology Planning \& Evaluation(IITP) grant funded by the Korea government(MSIT)
	(No.2022-0-00420,Development of Core Technologies enabling 6G End-to-End On-Time Networking \& No.2021-0-00270, Development of 5G MEC framework to improve food factory productivity, automate and optimize flexible packaging)

	\bibliographystyle{ieeetr}
	\bibliography{./VTC_2023_SplitAMC}
	
\end{document}